\documentstyle[prl,aps,graphicx,multicol]{revtex}
\newif\ifmynarrow \mynarrowfalse
\renewcommand{\narrowtext}{%
  \ifmynarrow\hspace*{\fill}\raisebox{-1ex}[0pt][0pt]{%
    \rule{0.3pt}{1ex}%
    \rule[1ex]{20.5pc}{0.3pt}}\fi
  \mynarrowtrue
  \vspace{-1.0ex}%
  \begin{multicols}{2}%
  \par\global\columnwidth20.5pc
  \global\hsize\columnwidth\global\linewidth\columnwidth
  \global\displaywidth\columnwidth}

\renewcommand{\widetext}{%
  \end{multicols}%
  \vspace{-2.5ex}%
  \noindent\raisebox{1ex}[0pt][0pt]{%
    \rule{20.5pc}{0.3pt}%
    \rule{0.3pt}{1ex}}%
  \par\global\columnwidth42.5pc
  \global\hsize\columnwidth\global\linewidth\columnwidth
  \global\displaywidth\columnwidth}

\begin{document}

\title{Highly anisotropic $g$-factor of two-dimensional hole systems}

\draft

\author{R. Winkler}
\address{Institut f\"ur Technische Physik III, Universit\"at
Erlangen-N\"urnberg, Staudtstr. 7, D-91058 Erlangen, Germany}

\author{S. J. Papadakis, E. P. De Poortere, and M. Shayegan}
\address{Department of Electrical Engineering, Princeton University,
Princeton, New Jersey 08544}

\date{May 28, 2000}
\maketitle
\begin{abstract}
  Coupling the spin degree of freedom to the anisotropic orbital
  motion of two-dimensional (2D) hole systems gives rise to a highly
  anisotropic Zeeman splitting with respect to different
  orientations of an in-plane magnetic field $B$ relative to the
  crystal axes.  This mechanism has no analogue in the bulk band
  structure.  We obtain good, qualitative agreement between theory
  and experimental data, taken in GaAs 2D hole systems grown on
  (113) substrates, showing the anisotropic depopulation of the
  upper spin subband as a function of in-plane $B$.
\end{abstract}

\narrowtext

Since the early days of two-dimensional (2D) carrier systems in
semiconductors it has been commonly assumed that the Zeeman energy
splitting, ${\Delta}E = g^\ast \mu_B B$, with $g^\ast$ the effective
$g$-factor and $\mu_B$ the Bohr magneton, is independent of the
direction of the external magnetic field B \cite{fan68}. Recently,
however, calculations and experiments have shown that $g^\ast$ can
have different values for $B$ applied in the direction normal to the
plane of the 2D system compared to in-plane
\cite{ivc92,lin91,gol93,pey93,kuh94}.  Here we report
calculations and experimental data for 2D holes occupying the heavy
hole subband, demonstrating that, even for a purely in-plane $B$,
$g^\ast$ can depend strongly on the orientation of $B$ with respect
to the crystal axes.  We show that the mechanism giving rise to this
remarkable g-factor anisotropy is fundamentally different from the
one responsible for the in-plane/out of plane anisotropy of $g^\ast$
discussed in previous work
\cite{ivc92,lin91,gol93,pey93,kuh94}: the latter is basically
a consequence of the bulk Zeeman Hamiltonian whereas the in-plane
anisotropy we are reporting stems from the {\em combined} effect of
the bulk $g^\ast$ and the anisotropic orbital motion of 2D hole
states in semiconductor quantum wells.

\begin{figure}
\centerline{\includegraphics[width=0.6\columnwidth]{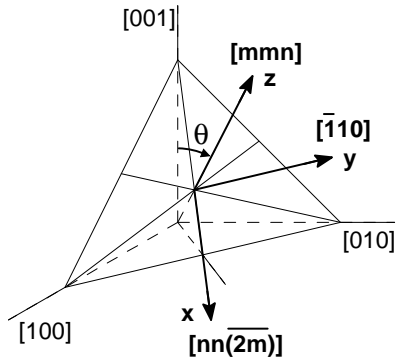}}
\vspace{3mm}
\caption[]{\label{fig_koord} Coordinate system for QW's grown
in $[mmn]$ direction ($z$ direction). Here $\theta$ is the angle
between $[mmn]$ and $[001]$, i.e., we have $\theta = \arccos (n /
\sqrt{2m^2+n^2})$. The axes for the in-plane motion are
$[nn\overline{(2m)}]$ ($x$) and $[\overline{1}10]$ ($y$).}
\end{figure}

In bulk semiconductors the motion of electrons and holes in the
presence of spin-orbit interaction gives rise to a $g^\ast$ which is
significantly modified compared to the free particle $g$-factor $g_0
= 2$ (Ref.\ \cite{rot59}). The resulting $g^\ast$ of holes (and
electrons) is nearly isotropic. Commonly, the isotropic part of the
hole $g^\ast$ is denoted by $\kappa$ \cite{lut56}. The anisotropic
part, $q$, is typically two orders of magnitude smaller than
$\kappa$ and, in the present discussion, is neglected completely.
The smallness of $q$ is in sharp contrast to the orbital motion of
holes for which we have highly anisotropic effective masses $m^\ast$
reflecting the spatial anisotropy of the crystal structure. In a 2D
hole system (2DHS) we have heavy hole (HH) subbands ($z$ component
of angular momentum $M=\pm 3/2$) and light hole (LH) subbands
($M=\pm1/2$). In the presence of an in-plane $B$, $\kappa$ couples
the two LH states, and the HH states to the LH states \cite{gol93}.
But there is no direct coupling between the HH states proportional
to $\kappa$. Therefore, the authors of Refs.\ 
\cite{lin91,gol93,pey93,kuh94} concluded that the Zeeman splitting
of HH states due to an in-plane $B$ is suppressed. However, this
result is correct only for quantum wells (QW's) grown in the
crystallographic high-symmetry directions [001] and [111] as for the
other growth directions we show that a new mechanism gives rise to a
large and highly anisotropic Zeeman splitting with respect to
different orientations of the in-plane magnetic field $B$ relative
to the crystal axes. In the following we will discuss QW's grown in
the crystallographic $[mmn]$ direction (with $m,n$ integers). For
this purpose we use the coordinate system shown in
Fig.~\ref{fig_koord} with $\theta$ denoting the angle between
$[mmn]$ and $[001]$. We remark that recently QW's for 2DHS's have
often been grown in [113] direction as this yields particularly high
hole mobilities \cite{dav91,pap99}.

We describe the hole subband states by means of the $4\times 4$
Luttinger Hamiltonian \cite{lut56}. For the in-plane ${\bf B} =
(B_x,B_y,0)$ we use the vector potential ${\bf A} = (zB_y,-zB_x,0)$.
Treating ${\bf A}$ and ${\bf B}$ by means of degenerate perturbation
theory we obtain in second order for $g^\ast$ at the bottom of the
HH subbands in an infinitely deep rectangular QW
\widetext
\begin{mathletters}
  \label{eq:gfak}
\begin{eqnarray}
  \label{eq:gfakx}
g^{\rm HH}_{[nn\overline{(2m)}]} & = &
6 \big[2 - 3 \sin^2 (\theta)\big] \sin (\theta)
\sqrt{4 - 3\sin^2 (\theta)} \; \frac{\kappa \, (\gamma_3 -
\gamma_2)}{\gamma_z^{\rm HH} - \gamma_z^{\rm LH}} \\
  \label{eq:gfaky}
g^{\rm HH}_{[\overline{1}10]} & = & -
6  \big[2 - 3 \sin^2 (\theta)\big] \sin^2 (\theta) \;
\frac{\kappa \, (\gamma_3 - \gamma_2)}
{\gamma_z^{\rm HH} - \gamma_z^{\rm LH}}
\end{eqnarray}
\end{mathletters}
\narrowtext\noindent
with
\begin{mathletters}
  \label{eq:pref}
\begin{eqnarray}
  \label{eq:prefhh}
\gamma_z^{\rm HH} & = & - \gamma_1 + 2 \big[(1-\alpha)
\gamma_2 + \alpha\gamma_3\big] \\
  \label{eq:preflh}
\gamma_z^{\rm LH} & = & - \gamma_1 - 2 \big[(1-\alpha)
\gamma_2 + \alpha\gamma_3\big] \\
  \label{eq:prefa}
\alpha & = & \sin^2 (\theta) \big[3 - {\textstyle\frac{9}{4}} \sin^2
(\theta) \big] .
\end{eqnarray}
\end{mathletters}\noindent
Here $\gamma_1$, $\gamma_2$, and $\gamma_3$, are the Luttinger
parameters \cite{lut56} and $\gamma_z^{\rm HH}$ and $\gamma_z^{\rm
LH}$ are the reciprocal effective masses in $z$-direction in the
axial approximation for the HH and LH subbands, respectively
\cite{delta}. In Eq.\ (\ref{eq:gfak}) the factor $(\gamma_3 -
\gamma_2)$ characterizes the anisotropy of the bulk valence band
\cite{lip70}. Hence, the anisotropic $g$-factor (\ref{eq:gfak}) is
due to the {\em combined} effect of the isotropic bulk $g$-factor
$\kappa$ and the anisotropy of the valence band. It disappears in
the axial limit $(\gamma_3 - \gamma_2) = 0$. The Zeeman splitting
discussed here has no analogue in the bulk band structure. It is a
unique feature of 2D systems. Therefore, Eq.\ (\ref{eq:gfak}) is
fundamentally different from the anisotropy of $g^\ast$ discussed in
Refs.\ \cite{ivc92,lin91,gol93,pey93,kuh94} as the latter
originates in properties of the bulk Zeeman Hamiltonian. Our sign
convention for $g^\ast$ used in Eq.\ (\ref{eq:gfak}) corresponds to
the dominant spinor component of the multicomponent eigenstates
(using a basis of angular momentum eigenfunctions with quantization
axis in the direction of $B$).
Note that in unstrained QW's the topmost subband is the HH1
subband.

For LH subbands in an in-plane $B$ as well as for HH and LH subbands
in a perpendicular $B$, $g^\ast$ contains terms similar to Eq.\ 
(\ref{eq:gfak}). However, the dominant contribution is given by the
bulk $g$-factor $\kappa$. For LH subbands in an in-plane $B$ we have
basically $g^{\rm LH}_\| = 4\kappa$, while for a perpendicular $B$
we have $g^{\rm HH}_z = 6\kappa$ and $g^{\rm LH}_z = 2\kappa$
(Refs.\ \cite{kuh94,lut56}). Yet, this implies a remarkable
difference \cite{lin91,gol93,pey93,kuh94} compared with Eq.\ 
(\ref{eq:gfak}).

\begin{figure}
\centerline{\includegraphics[width=0.7\columnwidth]{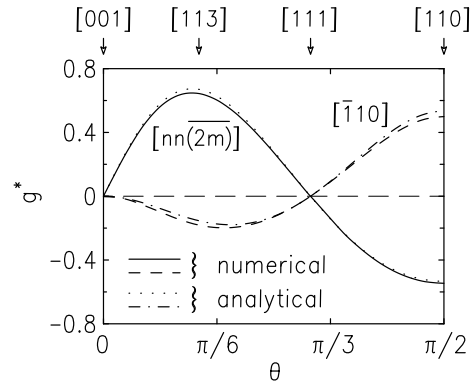}}
\vspace{3mm}
\caption[]{\label{pic:gaas} Anisotropic effective $g$-factor
$g^\ast$ of the HH1 subband for a 200 {\AA} wide
GaAs/Al$_{0.3}$Ga$_{0.7}$As QW as a function of $\theta$, the angle
between [001] and the growth direction. Results are shown for the
in-plane $B$ along the $[nn\overline{(2m)}]$ and $[\overline{1}10]$
directions. The solid and dashed lines were obtained by means of a
numerical diagonalization of the Luttinger Hamiltonian. The dotted
and dashed-dotted lines were obtained by means of Eq.\ 
(\ref{eq:gfak}).}
\end{figure}

In Fig.\ \ref{pic:gaas} we show the anisotropic $g^\ast$ of the HH1
subband for a 200 {\AA} wide GaAs/Al$_{0.3}$Ga$_{0.7}$As QW as a
function of the angle $\theta$. The analytical expressions
(\ref{eq:gfak}) (dotted and dashed-dotted lines) are in very good
agreement with the more accurate results obtained by means of a
numerical diagonalization \cite{gol93,win93} of the Luttinger
Hamiltonian (solid and dashed lines). Figure \ref{pic:gaas}
demonstrates that $g^\ast$ can be very anisotropic \cite{perp}. For
example, for the growth direction [113], $g^\ast$ is about a factor
of 4 larger when $B \parallel [33\overline{2}]$ compared to when $B
\parallel [\overline{1}10]$. Moreover, the sign of $g^{\rm
HH}_{[nn\overline{(2m)}]}$ is opposite to the sign of $g^{\rm
HH}_{[\overline{1}10]}$.

Equation (\ref{eq:gfak}) is applicable to a wide range of cubic
semiconductors with results qualitatively very similar to Fig.\ 
\ref{pic:gaas}. In particular, the relative anisotropy
\begin{equation}
\label{eq:anis}
\frac{g^{\rm HH}_{[\overline{1}10]}}
{g^{\rm HH}_{[nn\overline{(2m)}]}} = 
- \frac{\sin (\theta)}{\sqrt{4 - 3\sin^2 (\theta)}}
\end{equation}
is independent of the material-specific parameters $\gamma_i$ and
$\kappa$. This remarkable result can be traced back to the fact that
the anisotropy for different directions $\theta$ in ${\bf k}$-space
is always characterized by the single parameter $(\gamma_3 -
\gamma_2)$ (Ref.\ \cite{lip70}). Note that for QW's based on
narrow-gap semiconductors we have a larger $\kappa$ and smaller
effective masses so that for these materials the absolute values of
$g^\ast$ are significantly larger than $g^\ast$ of GaAs shown in
Fig.\ \ref{pic:gaas}, but the $g^\ast$ anisotropy is still given by
Eq.\ (\ref{eq:anis}) and depends only on $\theta$. The Zeeman
splitting can be even further enhanced if one uses semimagnetic
semiconductors containing, e.g., Mn. For these materials the
structure of the Hamiltonian is identical to the conventional
Luttinger Hamiltonian in the presence of a magnetic field with
$\kappa$ replaced by the effective $g$-factor due to the
paramagnetic exchange interaction \cite{pey93,kuh94}. Therefore Eq.\ 
(\ref{eq:gfak}) and the $g^\ast$ anisotropy [Eq.\ (\ref{eq:anis})]
are readily applicable to semimagnetic materials, also.

In a Taylor expansion of the Zeeman splitting, $\Delta E (B)$,
$g^\ast$ (times $\mu_B$) is the prefactor for the lowest order term
linear in $B$. Often terms of higher order in $B$ are neglected
because of their relevant insignificance. An interesting feature of
Fig.\ \ref{pic:gaas} is the vanishing of $g^\ast$ for the
high-symmetry growth directions [001] and [111] (Ref.\ 
\cite{gol93}). For the 2DHS discussed in Fig.\ \ref{pic:gaas} this
results in a splitting $\Delta E$ which at $B=1$~T is more than 2
orders of magnitude smaller than $\Delta E$ for growth directions
[113] and [110]. For these high-symmetry directions $\Delta E$ is
proportional to $B^3$ (Ref.\ \cite{mar90}). In second order
perturbation theory we obtain for the HH1 subband of an infinitely
deep rectangular QW of width $d$ grown in [001] direction
\widetext
\begin{equation}
  \label{eq:cub}
  \Delta E = (\mu_B B)^3 \left(\frac{m_0 d^2}{\pi^2\hbar^2}\right)^2
  \sqrt{\gamma_2^2 \cos^2(2\phi) + \gamma_3^2 \sin^2(2\phi)} 
  \left[\frac{4\kappa \, (\pi^2 - 6)}
        {\gamma_z^{\rm HH} - \gamma_z^{\rm LH}}
        + \frac{27 \gamma_3}{\gamma_z^{\rm HH} - 9\gamma_z^{\rm LH}}
        \right] \! .
\end{equation}
\narrowtext\noindent
Here $\phi$ is the angle between the in-plane ${\bf B}$ and the
[100] axis. The first term in the square bracket stems from ${\bf
k}\cdot {\bf p}$ coupling between the subbands HH1 and LH1, and the
second term is due to coupling between HH1 and LH3. We get similar,
though somewhat longer expressions for growth direction $[111]$. It
is remarkable that we have a nonzero $\Delta E$ even in the limit
$\kappa=0$. This can be understood as follows: The $4\times 4$
Luttinger Hamiltonian \cite{lut56} which is underlying our
calculations corresponds to an infinitely large spin-orbit splitting
between the topmost valence band $\Gamma_8^v$ and the split-off band
$\Gamma_7^v$. Therefore spin-orbit interaction is not explicitely
visible in our results, though, similar to Zeeman splitting in bulk
semiconductors \cite{rot59}, Eqs.\ (\ref{eq:gfak}) and
(\ref{eq:cub}) are a consequence of spin-orbit interaction. In 2D
systems the motion of electrons and holes in the presence of this
interaction can give rise to a Zeeman splitting even without a bulk
$g^\ast$. We remark that in a parabolic QW the in-plane $g^\ast$ of
the HH subbands also contains such terms independent of $\kappa$. We
have here a 2D analogue of Roth's famous formula \cite{rot59} for
the electron bulk $g^\ast$. Finally we note that, unlike
\cite{delta} Eq.\ (\ref{eq:gfak}), $\Delta E$ in Eq.\ (\ref{eq:cub})
increases proportional to $d^4$, i.e., Zeeman splitting is most
efficiently suppressed in narrow QW's.

Now we will show that the anisotropy of $g^\ast$ can be probed
experimentally by measuring the magnetoresistance of a high-mobility
2DHS as a function of in-plane $B$. The samples are 200 {\AA} wide
Si-modulation doped GaAs QW's grown on (113)A GaAs substrates. These
samples exhibit a mobility anisotropy believed to be due to an
anisotropic surface morphology \cite{her94}. They are patterned with
an L-shaped Hall bar to allow simultaneous measurements of the
resistivity along the $[33\overline{2}]$ and $[\overline{1}10]$
directions. Front and back gates are used to control the 2D density
in the QW and the perpendicular electric field which characterizes
the asymmetry of the sample \cite{pap99}.

\begin{figure}
\centerline{\includegraphics[width=0.95\columnwidth]{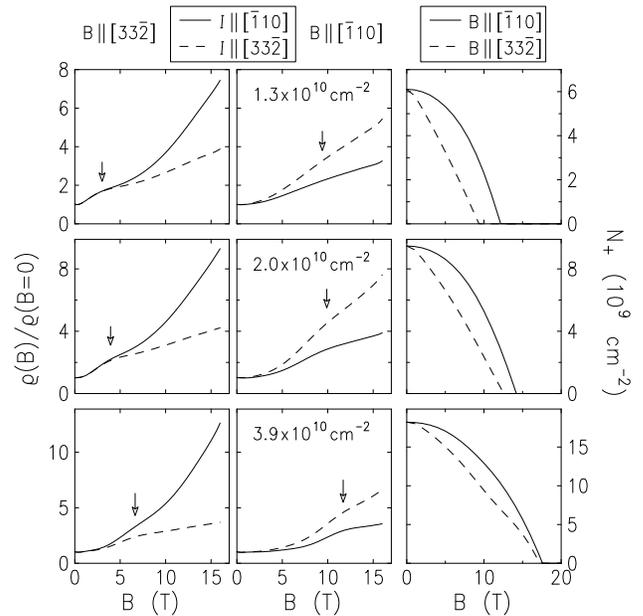}}
\vspace{3mm}
\caption[]{\label{pic:exp}Left and central panels:
Fractional change in resistivity $\rho(B)/\rho(0)$ due to an
in-plane $B$, measured at $T=0.3$~K in a GaAs 2D hole system grown
on a (113) substrate, for different 2D densities as indicated. The
arrows mark $B^\ast$ as defined in the text. Right panel: Calculated
density $N_+$ in the upper spin subband as a function of $B$.}
\end{figure}

The left two panels of Fig.\ \ref{pic:exp} show the resistivity
$\rho$ measured as a function of in-plane $B$ for three different
densities and different relative orientations of $B$ and current
$I$. For easier comparison we have plotted the fractional change
$\rho (B) / \rho(B=0)$. Apart from an overall positive
magnetoresistance these curves show a broad feature consisting of an
inflection point followed by a reduction in slope followed by
another inflection point. In Fig.\ \ref{pic:exp} we have placed
arrows between the two inflection points at a value of $B$ we call
$B^\ast$. Similar, though sharper features have been observed in
systems with several occupied confinement subbands when a subband is
depopulated by means of an in-plane $B$ \cite{jo93}.  We propose
that the magnetoresistance feature at $B^\ast$ in Fig.\ 
\ref{pic:exp} is related to a spin-subband depopulation and the
resulting changes in subband mobility and intersubband scattering as
the in-plane $B$ is increased.  Note that in each panel of Fig.\ 
\ref{pic:exp}, $B^\ast$ is the same for both current directions even
though the magnetoresistance is very different.  This implies that
$B^\ast$ depends on parameters which do not depend on current
direction.  This supports our hypothesis, as spin-subband
depopulation should not depend on the direction of current in the
sample.  

Our interpretation of $B^\ast$ is obviously consistent with $B^\ast$
in Fig.\ \ref{pic:exp} becoming larger with increasing density. It
is remarkable that $B^\ast$ for the $B \parallel [33\overline{2}]$
traces is about 4~T smaller than for the $[\overline{1}10]$ traces,
regardless of the $I$ direction. We associate this with the
anisotropy of the in-plane $g^\ast$. This interpretation is
validated by our self-consistently calculated
\cite{gol93,delta,win93} results for the density $N_+$ of the upper
spin subband as a function of $B$, shown in the right panel of Fig.\ 
\ref{pic:exp}. The density $N_+$ decreases much faster for $B
\parallel [33\overline{2}]$ than for $B \parallel [\overline{1}10]$,
in agreement with Fig.\ \ref{pic:gaas}.  We have further support for
our interpretation of $B^\ast$ from experiments where we increase
the asymmetry of the confining potential by means of the front and
back gates while keeping the 2D density fixed. We observe an increase
in $B^\ast$, in agreement with the results of our self-consistent
calculations.

One might ask whether the data in Fig.\ \ref{pic:exp} actually could
be summarized by a single value of $g^\ast$ for each trace.
Unfortunately, this is not possible because, due to the complicated
band structure of holes, $g^\ast$ depends on energy $E$, and we are
averaging over $g^\ast(E)$ for $E$ between the subband edge and the
Fermi energy. The importance of this effect can be readily deduced
from the right panel of Fig.\ \ref{pic:exp}, as we would have
straight lines for $N_+ (B)$ if $g^\ast$ (and the effective mass
$m^\ast$) were not dependent on $E$.

In Fig.\ \ref{pic:exp} the measured $B^\ast$ is significantly
smaller than the calculated $B$ for a complete depopulation of the
upper spin-subband. We note that for our low-density samples it can
be expected that $g^\ast$ is enhanced due to the exchange
interaction \cite{fan68,oka99}. This effect was not taken into
account in our self-consistent calculations. The qualitative
agreement between the experimental data and the calculations,
however, implies that these many-particle effects do not affect the
anisotropy of $g^\ast$.

The large anisotropy of the Zeeman splitting in 2DHS's offers many
possible device applications. In a polycrystalline material, e.g.,
one could alter the degree of spin polarization in different domains
by changing the direction of the external $B$. As can be seen in
Fig.\ \ref{pic:gaas}, because of the sign reversal of $g^\ast$ it is
even possible to have different domains with opposite spin
polarization for a given direction of $B$. Recently, there has been
a growing interest in controlling the spin degree of freedom for
quantum computing and spin electronics. In Ref.\ \onlinecite{div99}
the authors have sketched a quantum device which makes use of the
spatial variation of $g^\ast$ in layered semiconductor structures
made of, e.g., Al$_x$In$_y$Ga$_{1-x-y}$As. However, the authors have
estimated that a substantial change in $g^\ast$ requires a fairly
large electric field of the order of 100~kV/cm. Oestreich {\em et
al.} \cite{oes99} and Fiederling {\em et al.} \cite{fie99} have
suggested a spin aligner based on semimagnetic semiconductors. Here
the $g$-factor anisotropy of 2DHS's provides a powerful additional
degree of freedom for engineering such devices.

In conclusion, we have shown that coupling the spin degree of
freedom to the anisotropic orbital motion of 2D hole systems gives
rise to a highly anisotropic Zeeman splitting with respect to
different orientations of an in-plane magnetic field relative to the
crystal axes.

R.\ W.\ wants to thank O.\ Pankratov and P.\ T.\ Coleridge for
stimulating discussions and suggestions. Work at Princeton
University was supported by the NSF and ARO.



\widetext

\begin{thebibliography}{99}
\vspace{-10mm}

\bibitem{fan68} F.\ F.\ Fang and P.\ J.\ Stiles,
 Phys.\ Rev.\ {\bf 174}, 823 (1968).

\bibitem{ivc92} E.\ L.\ Ivchenko and A.\ A.\ Kiselev, Sov.\ Phys.\
 Semicond.\ {\bf 26}, 827 (1992);
 E.\ L.\ Ivchenko, A.\ A.\ Kiselev, and M.\ Willander,
 Solid State Commun.\ {\bf 102}, 375 (1997). 
 %
 These authors have shown that nonparabolicity leads to a small
 difference between the electron $g^\ast$ for a perpendicular and
 for an in-plane $B$. An experimental confirmation of these findings
 was given, e.g., by B.\ Kowalski {\em et al.}
 [Phys.\ Rev.\ B {\bf 49}, 14786 (1994)].

\bibitem{lin91} S.\ Y.\ Lin {\em et al.},
 Phys.\ Rev.\ B {\bf 43}, 12110 (1991).

\bibitem{gol93} G.\ Goldoni and A.\ Fasolino,
 Phys.\ Rev.\ B {\bf 48}, 4948 (1993).

\bibitem{pey93} P.\ Peyla {\em et al.},
 Phys.\ Rev.\ B {\bf 47}, 3783 (1993).

\bibitem{kuh94} B.\ Kuhn-Heinrich {\em et al.},
 Solid State Commun.\ {\bf 91}, 413 (1994).

\bibitem{rot59} L.\ M.\ Roth, B.\ Lax, and S.\ Zwerdling,
 Phys.\ Rev.\ {\bf114}, 90 (1959).

\bibitem{lut56} J.\ M.\ Luttinger,
 Phys.\ Rev.\ {\bf 102}, 1030 (1956).

\bibitem{dav91} A.\ G.\ Davies {\em et al.},
 J.\ Crystal Growth {\bf 111}, 318 (1991).

\bibitem{pap99} S.\ J.\ Papadakis {\em et al.},
 Science {\bf 283}, 2056 (1999).
 
\bibitem{delta} Eq.\ (\ref{eq:gfak}) is independent of the well
 width $d$. Similar to optical anisotropy in QW's [R.\ Winkler and
 A.\ I.\ Nesvizhskii, Phys.\ Rev.\ B {\bf 53}, 9984 (1996); and
 references therein] $g^\ast$ for an in-plane $B$ will depend on $d$
 if the split-off valence band $\Gamma_7^v$ is explicitely taken
 into account. Here, we omit the rather lengthy modifications in
 Eq.\ (\ref{eq:gfak}).  They become relevant for narrow QW's with
 $d\lesssim 50$ {\AA}. The traces in the right panel of Fig.\ 
 \ref{pic:exp} were obtained by means of a more complete $8\times 8$
 Hamiltonian containing the bands $\Gamma_6^c$, $\Gamma_8^v$ and
 $\Gamma_7^v$. However, the results obtained by means of the
 $4\times 4$ Luttinger Hamiltonian are similar.

\bibitem{lip70} N.\ O.\ Lipari and A.\ Baldereschi,
 Phys.\ Rev.\ Lett.\ {\bf 25}, 1660 (1970).

\bibitem{win93} R.\ Winkler and U.\ R\"ossler,
 Phys.\ Rev.\ B {\bf 48}, 8918 (1993).

\bibitem{perp} For comaprison, we remark that for the GaAs system in
 Fig.\ \ref{pic:gaas} we have $g_z^{\rm HH} = 6 \kappa \approx 7.2$.
 
\bibitem{mar90} R.\ W.\ Martin {\em et al.}  [Phys.\ Rev.\ B {\bf
 42}, 9237 (1990)] have predicted a Zeeman splitting for growth
 direction [001] proportional to $B^2$. This result was obtained by
 an incorrect symmetrization of the $k$ dependent terms in the
 Luttinger Hamiltonian. The cubic dependence of $\Delta E$ on $B$ in
 the approximate, analytical expression (\ref{eq:cub}) is in very
 good agreement with our more accurate numerical calculations based
 on Refs.\ \onlinecite{gol93,win93}.

\bibitem{her94} J.\ J.\ Heremans {\em et al.},
 J.\ Appl.\ Phys.\ {\bf 76}, 1980 (1994).

\bibitem{jo93} J.\ Jo {\em et al.},
 Phys.\ Rev.\ B {\bf 47}, 4056 (1993).

\bibitem{oka99} T.\ Okamoto {\em et al.},
 Phys.\ Rev.\ Lett.\ {\bf 82}, 3875 (1999).
 
\bibitem{div99} D.\ P.\ DiVincenzo {\em et al.}, in {\em Quantum
 Mesoscopic Phenomena and Mesoscopic Devices in Microelectronics},
 edited by I.\ O.\ Kulik and R.\ Ellialtioglu, (NATO Advanced Study
 Institute, 1999, cond-mat/9911245).

\bibitem{oes99} M.\ Oestreich {\em et al.},
 Appl.\ Phys.\ Lett.\ {\bf 74}, 1251 (1999).

\bibitem{fie99} R.\ Fiederling {\em et al.},
 Nature {\bf 402}, 787 (1999).

\end{thebibliography}
\end{document}